\documentclass[aps,twocolumn]{revtex4-2}

\usepackage{amsmath,amssymb,amsfonts,color,graphicx,tabularx}

\usepackage[unicode=true,colorlinks=true]{hyperref}

\hypersetup{linkcolor=blue,citecolor=blue,urlcolor=blue}

\renewcommand{\figurename}{\textbf{Fig.}}

\begin{document}

\title{Tunable even- and odd-denominator fractional quantum Hall states \\ in trilayer graphene}

\author{Yiwei Chen$^{1}$$^{\dagger}$}
\author{Yan Huang$^{1}$$^{\dagger}$}
\author{Qingxin Li$^{1}$$^{\dagger}$}
\author{Bingbing Tong$^{2,3}$}
\author{Guangli Kuang$^{4}$}
\author{Chuanying Xi$^{4}$}
\author{Kenji Watanabe$^{5}$}
\author{Takashi Taniguchi$^{6}$}
\author{Guangtong Liu$^{2,3,7\ast}$}
\author{Zheng Zhu$^{8}$}
\author{Li Lu$^{2,3,7}$}
\author{Fu-Chun Zhang$^{8,9,10}$}
\author{Ying-Hai Wu$^{11 \ast}$}
\author{Lei Wang$^{1,10 \ast}$}

\affiliation{$^{1}$National Laboratory of Solid-State Microstructures, School of Physics, Nanjing University, Nanjing, 210093, China}
\affiliation{$^{2}$Beijing National Laboratory for Condensed Matter Physics and Institute of Physics, Chinese Academy of Sciences, Beijing 100190, China}
\affiliation{$^{3}$Hefei National Laboratory, Hefei 230088, China.}
\affiliation{$^{4}$Anhui Province Key Laboratory of Condensed Matter Physics at Extreme Conditions, High Magnetic Field Laboratory of the Chinese Academy of Science, Hefei 230031, China.}
\affiliation{$^{5}$Research Center for Electronic and Optical Materials, National Institute for Materials Science, 1-1 Namiki, Tsukuba 305-0044, Japan}
\affiliation{$^{6}$Research Center for Materials Nanoarchitectonics, National Institute for Materials Science, 1-1 Namiki, Tsukuba 305-0044, Japan}
\affiliation{$^{7}$Songshan Lake Materials Laboratory, Dongguan 523808, China}
\affiliation{$^{8}$Kavli Institute of Theoretical Sciences, University of Chinese Academy of Sciences, Beijing, 100049, China}
\affiliation{$^{9}$CAS Center for Excellence in Topological Quantum Computation, University of Chinese Academy of Sciences, Beijing, 100049, China}
\affiliation{$^{10}$Collaborative Innovation Center of Advanced Microstructures, Nanjing University, Nanjing, 210093, China}
\affiliation{$^{11}$School of Physics and Wuhan National High Magnetic Field Center, Huazhong University of Science and Technology, Wuhan 430074, China}

\affiliation{$^{\dagger}$These authors contributed equally to this work.}
\affiliation{$^{\ast}$Corresponding authors, Email: gtliu@iphy.ac.cn; yinghaiwu88@hust.edu.cn; leiwang@nju.edu.cn}


\maketitle
\textbf{The fractional quantum Hall (FQH) states are exotic quantum many-body phases whose elementary charged excitations are neither bosons nor fermions but anyons, obeying fractional braiding statistics. While most FQH states are believed to have Abelian anyons, the Moore-Read type states with even denominators, appearing at half filling of a Landau level (LL), are predicted to possess non-Abelian excitations with appealing potentials in topological quantum computation. These states, however, depend sensitively on the orbital contents of the single-particle LL wavefunction and the mixing between different LLs. Although they have been observed in a few materials, their non-Abelian statistics still awaits experimental confirmation. Here we show magnetotransport measurements on Bernal-stacked trilayer graphene (TLG), whose unique multiband structure facilitates the interlaced LL mixing, which can be controlled by external magnetic and displacement fields. We observe a series of robust FQH states including even-denominator ones at filling factors $\nu=-9/2$, $-3/2$, $3/2$ and $9/2$. In addition, we are able to finetune the LL mixing and crossings to drive quantum phase transitions of these half-filling states and their neighboring odd-denominator ones, exhibiting a related emerging and waning behavior. Our results establish TLG as a controllable system for tuning the weights of LL orbitals and mixing strength, and a fresh platform to seek for non-Abelian quasi-particles.}

Electrons confined in a two-dimensional system under a perpendicular magnetic field develop quantized energy levels. And fully filling such Landau levels (LLs) one by one gives rise to the integer quantum Hall states~\cite{Klitzing1980New}. Within a LL, the strong Coulomb interaction dominates over kinetic energy in the highly degenerated LL flatband, further conduces to the emergence of FQH states at certain fractional fillings $\nu$~\cite{tsui1982}. In most cases, the denominators of $\nu$ are odd integers, which finds an explanation in that FQH states can be understood effectively as integer quantum Hall states of composite fermions~\cite{jain1989composite}. An important exception to the odd-denominator rule is the 5/2 FQH state found in the second LL of GaAs~\cite{willett1987observation}. Based on extensive experimental and theoretical investigations~\cite{halperin2020,radu2008charge,venka2011charge,banerjee2018observation,dutta2022}, the most probable explanation of this state is the Moore-Read theory~\cite{moore1991nonabelions} and its extensions, which provide us the Pfaffian, anti-Pfaffian, and particle-hole symmetric Pfaffian wave functions as candidates~\cite{lee2007particle,levin2007particle,sondt2015,zucker2016}. The elementary charged excitations of these states obey non-Abelian braiding statistics and may be utilized to perform fault-tolerant quantum computation that are topologically protected at the fundamental level~\cite{nayak2008non}. In recent years, even-denominator FQH states have been observed in several other systems and some of them are also believed to host non-Abelian anyons~\cite{falson2015even,zibrov2017tunable,li2017even,YoungwookEven,huang2022valley,shi2020odd,hossain2023valley}.

\begin{figure*}[t!]
\begin{center}
\includegraphics[width=1\linewidth]{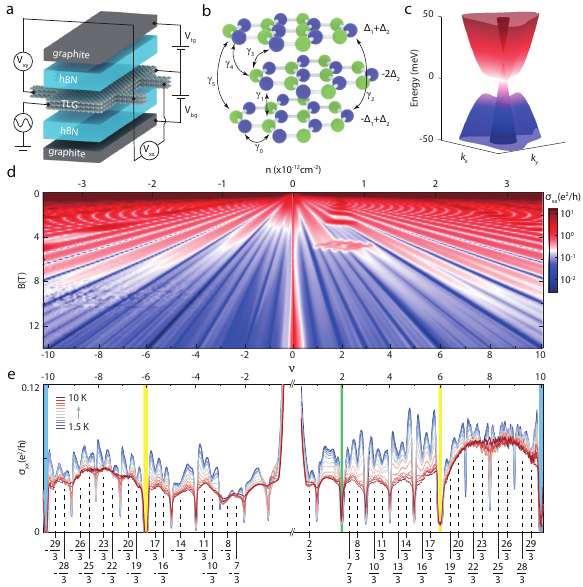}
\caption
{
\textbf{Band structure of TLG and its quantum Hall states.} 
        \textbf{a}, Schematic of the TLG device with graphite top-gate and bottom-gate isolated by hBN. 
        \textbf{b}, The crystal structure of Bernal stacked TLG and the parameters in its tight-binding model. 
        \textbf{c}, The low-energy band structure of TLG without the displacement field $D$ in the vicinity of the $\mathbf{K}_{+}$ valley.
        \textbf{d}, The color map of the longitudinal conductance $\sigma_{xx}$ plotted versus carrier density $n$ and magnetic field $B$ at 1.5 K and $D=0$ mV/nm. The filling factors defined at $B$ = 14 T is given below the bottom axis. The diamond pattern at $B \approx 5$ T is attributed to level crossings. A plethora of FQH states are observed above 10 T.
        \textbf{e}, $\sigma_{xx}$ as a function of $\nu$ for different temperatures at $B$ = 14 T and $D$ = 0 mV/nm. The filling factors $\nu=2$, $\pm 6$ and $\pm 10$ are marked by green, yellow, and blue shaded regions, respectively. In the range $-6{\leqslant}\nu{\leqslant}-6$, there are two MLG levels and four BLG levels for each spin component (see Fig.~\ref{fig:fig3}f).
}
\label{fig:fig1}
\end{center}
\end{figure*}

\begin{figure*}[t!]
\begin{center}
\includegraphics[width=1\linewidth]{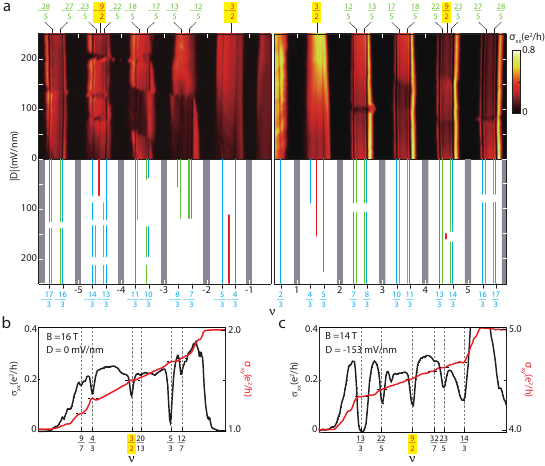}
\caption
{
\textbf{Even- and odd-denominator FQH states in the filling factor range $\boldsymbol{-6{\leqslant}\nu{\leqslant}6}$.} 
    \textbf{a}, Top panel: the color map of the longitudinal conductance $\sigma_{xx}$ plotted versus $\nu$ and the displacement field $D$. The magnetic field is $B=14$ T and the base temperature is 15 mK. FQH states with both even and odd denominators are observed at various filling factors $\nu=p/q$ ($p$ and $q$ are integers) as signified by minima of $\sigma_{xx}$. The FQH states exhibit intricate evolution with $D$. Bottom panel: sketch map of the top panel in which the FQH states are marked for clarity. Grey shaded regions label the integer quantum Hall states. Red, blue, green lines label the FQH states with $q=2$, $3$, and $5$, respectively. 
     \textbf{b-c}, The line plots of $\sigma_{xx}$ and the Hall conductance $\sigma_{xy}$ for $1{\leqslant}\nu{\leqslant}2$ with $B=16$ T and $D=0$ mV/nm (\textbf{b}) and $4{\leqslant}\nu{\leqslant}5$ with $B=14$ T, $D=-153$ mV/nm (\textbf{c}). A few FQH states are indicated by the dashed lines and the associated values of $\nu$ are given below. The short horizontal lines mark the quantized plateau of $\sigma_{xy}$.
}
\label{fig:fig2}
\end{center}
\end{figure*}

We focus on Bernal stacked trilayer graphene (TLG) here and report even-denominator FQH states in this system for the first time as well as a plethora of odd-denominator ones. In recent years, extensive investigations of monolayer and bilayer graphene (MLG, BLG)~\cite{zibrov2017tunable,li2017even,zibrov2018even,YoungwookEven,huang2022valley}, and experimental attempts have also been made on TLG. However, FQH states in TLG remained elusive with only tenuous traces speculated~\cite{CNLau2016tunable}. Compared to MLG and BLG, TLG possesses richer and more delicate band structure tunability~\cite{serbyn2013new}. Under zero displacement field, the band structure of TLG can be decomposed to a combination of MLG and BLG (with hopping parameters that are different from the actual monolayer and bilayer systems), but this fact by no means implies that the physics of TLG is a trivial repetition of MLG and BLG. In the presence of a magnetic field, the LLs originate from the MLG and BLG parts are not separated in energy but intersect with each other. If a vertical displacement field is introduced, the decomposition is no longer valid as the MLG and BLG parts hybridize. Each single-particle eigenstate is a superposition of the solutions in the non-relativistic (NR) Landau problem and its weights in different NR levels vary with external fields. The separations between LLs can be tuned to generate many different orderings. For GaAs, the NR second LL that hosts the $5/2$ state is sandwiched between the lowest and third LLs. In contrast, one level in TLG that is similar to the NR second LL may be surrounded from above and blow by other levels that have various different orbital contents. The LL mixing between them is very sophisticated and may results in intricate competition between strongly correlated states.

The structure of our TLG devices is depicted in Fig.~\ref{fig:fig1}a, where two graphite gates are separated from TLG by insulating hBN (see Extended Data Fig.~1 for optical images of our device). By applying voltages $V_{tg}$ on the top gate and $V_{bg}$ on the bottom gate, the carrier density $n$ and the displacement field $D$ can be tuned independently as: $n=(C_{b}V_{bg}+C_{t}V_{tg})/e$ and $D=(C_{b}V_{bg}-C_{t}V_{tg})/2$, where $C_{b}$, $C_{t}$ are average geometric capacitances for the bottom and top gates. The lattice structure of TLG is shown in Fig.~\ref{fig:fig1}b together with the Slonczewski–Weiss–McClure (SWMc) parameters in its tight-binding description~\cite{koshino2011landau}. The potential difference between the top and bottom layers caused by $D$ is denoted as $2\Delta_{1}$. An additional variable $\Delta_{2}$ was proposed to characterize the intrinsic charge imbalance between the outer and middle layers~\cite{serbyn2013new}. The low-energy band of TLG for $D=0$ is shown in Fig.~\ref{fig:fig1}c, where MLG-like linear and BLG-like quadratic components can be discerned. When a perpendicular magnetic field $B$ is applied, the linear and quadratic bands give rise to two sets of LLs that scale as $\sqrt{B}$ and $B$, respectively. Some levels from these two sets may intersect with each other when $B$ is varied. This can be seen in Fig.~\ref{fig:fig1}d that presents the Landau fan diagram measured at $1.5$ K. A few level crossings are observed around $B$ = 5 T for $2{\leqslant}\nu{\leqslant}6$, as was also reported in previous works~\cite{Pablo2011quantum,CNLau2016tunable,Deshmukh2017strong}. When the magnetic field increase to 11 T, minimal longitudinal conductance $\sigma_{xx}$ was observed at all integer filling factors $-10{\leqslant}\nu{\leqslant}10$, which signifies a complete lifting of the spin and valley degeneracies of the LLs by electron correlation. Meanwhile, a variety of well-developed FQH states emerge. The temperature dependence of $\sigma_{xx}$ at 14 T are displayed in Fig.~\ref{fig:fig1}e for many FQH states with denominator $3$, whose thermal activation behaviour clearly demonstrates their incompressible nature. An interesting feature is the different onset magnetic field for some states in the range $6{\leqslant}\nu{\leqslant}10$. FQH states at $\nu=19/3, 23/3, 25/3, 29/3$ were observed even {\it before} integer states appear at $\nu=7,9$ (see Extended Data Fig.~2 for $\sigma_{xx}$ in the $\nu-B$ plane). In contrast, FQH states begin to develop at $\nu=20/3,22/3,26/3,28/3$ near 10 T, at which point LL degeneracies at $\nu=7,9$ has already been lifted. This phenomenon suggests that some FQH states are intimately connected with the lifting of spin-valley degeneracy at $\nu=7,9$.

Next we investigate the states with $-6{\leqslant}\nu{\leqslant}6$ at lower temperatures in detail. The color map of $\sigma_{xx}$ versus $\nu$ and $D$ is plotted in Fig.~\ref{fig:fig2}a with the left (right) panel of the top row showing the hole (electron) side. The FQH states are sketched in the bottom row for clarity (see Extended Data Fig.~3 for the complete map). For various filling factors of the form $\widetilde{\nu}=\nu-[\nu]=s/(2s+1)$ and $1-s/(2s+1)$ ($[\nu]$ is the greatest integer less than or equal to $\nu$ and $s=1,2$), the observed minima of $\sigma_{xx}$ suggest the existence of FQH states. These states are illustrated as blue and green lines in the sketch panel. Besides these odd-denominator states, even-denominator FQH states were found at $\nu=-9/2,-3/2,3/2,9/2$ and highlighted by red lines. The $\nu=-9/2$ and $3/2$ states can be realized at $D=0$ and remain stable over a wide range of $D$, but the $\nu=-3/2$ and $9/2$ states only appears when a finite $D$ is applied. In fact, the $\nu=9/2$ state can only be observed in a very narrow range of $D$. We plot $\sigma_{xx}$ together with the Hall conductance $\sigma_{xy}$ around $\nu=3/2$ in Fig.~\ref{fig:fig2}b and the same quantities around $\nu=9/2$ in Fig.~\ref{fig:fig2}c. The concomitant appearance of exponentially suppressed $\sigma_{xx}$ and quantized $\sigma_{xy}$ clearly demonstrate that FQH states are realized at many different $\nu$. Interestingly, a weak minima can be seen at $\nu=20/13$, which may be a composite fermion state or a Pfaffian daughter state~\cite{levin2009collective}. As shown in Extended Data Fig.~4, similar features are observed in the vicinity of $\nu=-9/2,-3/2$. In contrast, there is no signature of daughter states associated with $\nu=9/2$ but FQH states are observed clearly at $\widetilde{\nu}=2/5$ and 3/5.

\begin{figure*}[t!]
\begin{center}
\includegraphics[width=1\linewidth]{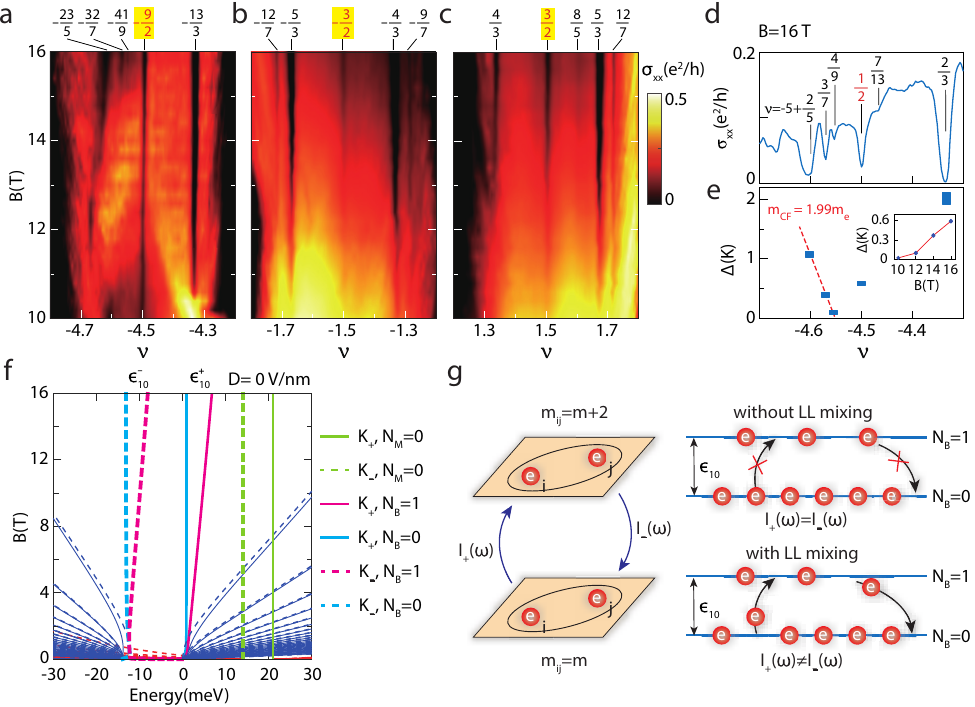}
\caption
{
\textbf{Evolution of FQH states with magnetic field.} 
    \textbf{a-c}, The color maps of the longitudinal conductance $\sigma_{xx}$ plotted versus $B$ and $\nu$ at 15 mK. The displacement field $D$ is zero in \textbf{a} and \textbf{c} and 217 mV/nm in \textbf{b}.
    \textbf{d}, $\sigma_{xx}$ as a function of $\nu$ at $B=16$ T. The vertical lines and the numbers mark some FQH states.
    \textbf{e}, The energy gaps of the FQH states marked in \textbf{d} deduced from thermal activation measurements. The data points are drawn as blue rectangles with heights proportional to their fitting uncertainty. The gaps for three odd-denominator states are fitted linearly using the red dash line $\Delta =\hbar e B_{\rm eff}/m_{\rm CF}$ ($B_{\rm eff}=(1-2\widetilde{\nu})B$ is the effective magnetic field for composite fermions, and $m_{\rm CF}$ is composite fermion mass). Its inset shows the gap at $\nu=-9/2$ for different $B$. 
    \textbf{f}, The theoretically computed LLs with $\Delta_{1}=\Delta_{2}=0$. The $\mathbf{K}_{+}$ ($\mathbf{K}_{-}$) valley is represented using solid (dashed) lines. The scheme of six levels is shown on the right of this panel. The MLG and BLG levels are marked by the subscripts ${\rm M/B}$ and their orbital contents are given by the numbers $0/1$. The splitting between the two BLG levels in the same valley is denoted as $\epsilon^{\pm}_{10}$.
    \textbf{g}, Schematic of the chiral graviton spectral functions $I_{\pm}(\omega)$. Left panel: Chiral gravitons are excited when the relative angular momentum of electron pairs $m$ is changed by $2$. $I_{+}(\omega) (I_{-}(\omega))$ is the spectral function of the operator that increases (decreases) $m$. Right panel: Particle-hole symmetry within one LL ensures that $I_{+}(\omega)=I_{-}(\omega)$, but LL mixing causes asymmetry between them. $\epsilon_{10}$ denotes the gap between the two LLs under consideration.
}
\label{fig:fig3}
\end{center}
\end{figure*}

For a fixed displacement field, the dependence of $\sigma_{xx}$ on $B$ around $\nu=-9/2,-3/2,3/2$ is presented in Fig.~\ref{fig:fig3}a,b,c. The existence of odd-denominator four-flux FQH states ($\nu=-32/7,-41/9,-12/7,-9/7,12/7$) underscores the high quality of our sample. The three even-denominator FQH states are robust in a considerable range of magnetic field. The line plot of $\sigma_{xx}$ at $B=16$ T with $-5{\leqslant}\nu{\leqslant}-4$ is shown in Fig.~\ref{fig:fig3}d on which a few FQH states are indicated. The energy gaps of some states in Fig.~\ref{fig:fig3}d are deduced from their thermal activated behaviour and presented in Fig.~\ref{fig:fig3}e (see Extended Data Fig.~5 for the fitting). For the odd-denominator ones at $\widetilde{\nu}=2/5,3/7,4/9$, the data can be understood using the composite fermion theory. A remarkable prediction of this theory is that the energy gap decreases to zero linearly as the filling factor $\widetilde{\nu}$ approaches 1/2~\cite{durr1993}. The gap values are fitted using $\Delta =\hbar e B_{\rm eff}/m_{\rm CF}$ in Fig.~\ref{fig:fig3}e, where $B_{\rm eff}=(1-2\widetilde{\nu})B$ is the effective magnetic field for composite fermions and $m_{\rm CF}=1.99m_{e}$ is the composite fermion mass. It is obvious that this rule is violated by the $\nu=-9/2$ FQH state whose energy gap is well above zero. The inset of Fig.~\ref{fig:fig3}e shows the evolution of the gap at $\nu=-9/2$ with $B$. It gets larger when $B$ increases as one would expect for a state driven by interaction.

\begin{figure*}[t!]
\begin{center}
\includegraphics[width=1\linewidth]{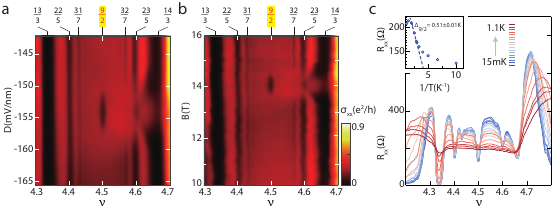}
\caption
{
\textbf{Displacement and magnetic fields tuning of FQH states in the filling factor range $\boldsymbol{4{\leqslant}\nu{\leqslant}5}$.} 
    \textbf{a-b}, The color map of longitudinal conductance $\sigma_{xx}$ plotted versus $D$ (with $B=14$ T fixed in \textbf{a}) or $B$ (with $D=-153$ mV/nm fixed in \textbf{b}) and $\nu$ at 15 mK. Phase transitions of FQH states are observed at $\nu=9/2,32/7,23/5$.
    \textbf{c}, The temperature dependence of $\sigma_{xx}$ versus $\nu$ for $D=-153$ mV/nm and $B=14$ T. Its inset shows the Arrhenius fitting of the thermal activation gap at $\nu=9/2$.
}
\label{fig:fig4}
\end{center}
\end{figure*}

To understand the even-denominator states, we first inspect the LLs of TLG presented in Fig.~\ref{fig:fig3}f (spin is suppressed for simplicity). The parameter $\Delta_{2}$ is fixed at zero in most parts of our discussion, but a small value would not be detrimental either. In the absence of $D$, the LLs can be labeled as MLG and BLG ones. For $B \; {\gtrsim} \; 8$ T and $-6{\leqslant}\nu{\leqslant}6$, there are three levels in each of the $\mathbf{K}_{\pm}$ valley and they are labeled using $N_{\rm M/B}=0,1$. The subscript ${\rm M/B}$ refers to their MLG or BLG origin and the numbers $0$ and $1$ characterize the orbital contents of the single-particle states. When the MLG and BLG parts are hybridized by a nonzero $D$, the single-particle eigenstates are six-dimensional vectors consist of NR Landau orbits. For one eigenstate of the TLG LL, the largest weight may reside in the the NR lowest LL, then it would be denoted as NR0. If there is a substantial weight in the NR second LL, it would be denoted as NR1. The $\mathbf{K}_{\pm},N_{\rm B}=0$ levels are of the NR0 type and the $\mathbf{K}_{\pm},N_{\rm B}=1$ levels are of the NR1 type. It is well-known that the NR second LL is favorable for realizing FQH state at half filling, as exemplified by the 5/2 state in GaAs~\cite{willett1987observation}. Using this information, we can provide a simple picture for the $\nu=-9/2$ and $3/2$ states at $D=0$. As indicated in Fig.~\ref{fig:fig3}f, electrons fill all the LLs below $\mathbf{K}_{-},N_{\rm B}=0$ at $\nu=-6$. An extra $3/2$ filling of electrons are added to arrive at $\nu=-9/2$. The Zeemann splitting is $E_{\rm Z} = 1.62$ meV at 14 T whereas the separation $\epsilon^{-}_{10}$ between $\mathbf{K}_{-},N_{\rm B}=1$ and $\mathbf{K}_{-},N_{\rm B}=0$ is $4.59$ meV. If there is no interaction, electrons would populate the $\mathbf{K}_{-},N_{\rm B}=0$ level with both spin projections. However, Coulomb interaction is likely to favor maximal spin polarization, so the electrons may instead occupy the $\mathbf{K}_{-},N_{\rm B}=0$ and $\mathbf{K}_{-},N_{\rm B}=1$ levels with spin down. This leads to a half-filled NR1 level in which the Moore-Read type states may be realized.

This picture for $\nu=-9/2$ is corroborated using exact diagonalization results on the torus. There are six quasi-degenerate ground states, which is consistent with the prediction for the Moore-Read type states as well as previous results in BLG~\cite{apalkov2011stable,papic2011tunable,snizhko2012,zhu2020widely}. This result alone cannot tell us if the state is of the Pfaffian, anti-Pfaffian, or particle-hole symmetric Pfaffian type. To this end, we have computed the chiral graviton spectral functions~\cite{liou2019,haldane2021}. As illustrated in Fig.~\ref{fig:fig3}g, these quantities are designed to reveal the relative angular momentum of electron pairs. It has been shown that the dominant chirality is negative (positive) for the Pfaffian (anti-Pfaffian) wave function~\cite{haldane2021}. If we only keep the $\mathbf{K}_{-},N_{\rm B}=1$ in our calculation, particle-hole symmetry ensures that the two chiralities are the same. After incorporating LL mixing with the $\mathbf{K}_{-},N_{\rm B}=0$ level (and excluding all other levels), the Pfaffian state becomes the favored one. This is consistent with the possible existence of a daughter state at $\nu=-5+7/13$. More details are given in the Supplementary Information. In general, when a NR0/NR1 doublet has $3/2$ filling, one may expect to see an even-denominator FQH state. For the $\nu=3/2$ state, the $\mathbf{K}_{+},N_{\rm B}=0$ and $\mathbf{K}_{+},N_{\rm B}=1$ levels are partially occupied. It should also be of the Pfaffian type in view of the weak minima at $\nu=20/13=1+7/13$. For the range of $B$ that have been studied, it may be sufficient to keep only the two BLG levels, but we should keep in mind that the MLG levels are not too far away in energy. If $B$ increases to about 29 T, the $\mathbf{K}_{+},N_{\rm B}=1$ level would be equidistant from $\mathbf{K}_{+},N_{\rm B}=0$ and $\mathbf{K}_{+},N_{\rm M}=0$. This results in exotic LL mixing that has not been explored in previous works and the fate of the $3/2$ state would be very interesting. A higher magnetic field is require to do the same thing for the $\mathbf{K}_{-}$ valley. 

When the displacement field is turned on, the FQH states evolves in different manners. The disappearance of the $\nu=-9/2$ and $3/2$ states can be explained qualitatively as shown in Extended Data Fig.~6b. The $\mathbf{K}_{-},N_{\rm B}=0$ level goes down when $D$ increases and eventually crosses with two NR2 type levels. In this regime, the $\nu=-9/2$ state would not correspond to a half-filled NR1 type level, so no even-denominator FQH state is expected. The experimental data in Fig.~\ref{fig:fig2}a indeed suggests that a level crossing occurs at $D \approx 80$ mV/nm. The $\mathbf{K}_{+},N_{\rm B}=0$ and $\mathbf{K}_{+},N_{\rm B}=1$ levels changes slowly with $D$, but they do cross with each other at quite large $D$. Another important change is that the weight of $\mathbf{K}_{+},N_{\rm B}=1$ in the NR second LL decreases. It is thus plausible that the $3/2$ state gradually weakens as $\epsilon^{+}_{10}$ decreases, which is consistent with the experimental data in Fig.~\ref{fig:fig2}a. In contrast to these two filling factors, a state emerges at $\nu=-3/2$ when $D$ becomes sufficiently large. For the case with $D=0$, the $\mathbf{K}_{+},N_{\rm B}=0$ level with spin down is half filled and no FQH state is expected. As $D$ increases, the $\mathbf{K}_{-},N_{\rm B}=1$ level goes up slowly and eventually crosses with $\mathbf{K}_{+},N_{\rm B}=0$ at large $D$. This crossing occurs at smaller $D$ if a small positive $\Delta_{2}$ is invoked (see Extended Data Fig.~6c). After the crossing, the electrons fully occupy the $\mathbf{K}_{\pm},N_{\rm B}=0$ levels with both spin components at $\nu=-2$. The $-3/2$ state would correspond to a half filled NR1 level and it is likely of the Pfaffian type given the weak feature observed at $\nu=-2+7/13$ (see Extended Data Fig.~4).

Finally, we study the $\nu=9/2$ FQH state that only appears in a very small window of $D$ and $B$. The color map of $\sigma_{xx}$ around $\nu=9/2$ is presented in Fig.~\ref{fig:fig4}a,b (with one parameter fixed and the other varied). As the $\nu=9/2$ state emerges, the gap at $\nu=23/5$ closes and then reopens. This implies that a phase transition has occurred between two different $\nu=23/5$ states. The $\nu=32/7$ state simply disappears when the $\nu=9/2$ state is observed. On the contrary, the states at $\nu=22/5$ and $31/7$ remain stable and no transition is found. By fitting the thermal activation data in Fig.~\ref{fig:fig4}c, the gap at $\nu=9/2$ is found to be $\sim 0.51$ K. For our tight-binding model at $\Delta_{2}=0$, the active levels at $\nu=9/2$ are of the NR0 type, which is unfavorable for realizing even-denominator states. If we change $\Delta_{2}$ to -10 meV, the Pfaffian state would be realized at $\nu=9/2$. This analysis is not very satisfactory because such a value of $\Delta_{2}$ is not quite reasonable and it is difficult to explain why this state only appear in a narrow range of $D$. To this end, we may consider multi-component FQH states whose spin and/or valley indices are not polarized. The Halperin 331 state is a well-known two-component state with $\nu=1/2$~\cite{halperin1983}. Another example is the Jain state constructed from the parton theory~\cite{jain1989parton,moran2012}. For a suitable range of $D$, two NR0 type levels are almost degenerate. At $\nu=9/2$, two NR0 type levels with the same valley index are half filled. The interaction between different valleys may be altered by valley anisotropic terms~\cite{kharitonov2012} such that the Jain state is stabilized~\cite{wu2022two}. This mechanism was proposed to explain the $1/2$ state observed in MLG~\cite{zibrov2018even}, and a more detailed analysis is needed to check if it also works for TLG.

In summary, our results unveil the rich odd- and even- denominator FQH states in TLG and underscore the extraordinary tunability due to intricate interplay of spin, valley, and orbital degrees of freedom. While the odd-denominator states are most acceptably described by the composite fermion theory, other candidates such as the Read-Rezayi states with non-Abelian Fibonacci anyons~\cite{readrezayi1999} are also possible and deserve further investigations. On the other hand, it would be fruitful to further explore the consequences brought by the evolution of LL eigenstates and their mixing and crossings. By varying external fields, we may switch between multiple Abelian and non-Abelian FQH states as well as non-FQH states, and continuous quantum phase transitions may be feasible in some of these process. Since FQH states are not described by the Landau paradigm based on symmetry breaking, their transitions are most likely not captured by the standard Landau-Ginzburg-Wilson theory~\cite{sachdevbook}. The low-energy effective theory of FQH states generally involve Chern-Simons gauge fields~\cite{XGWen1995Topological}, so there could be many exotic transitions described by strongly coupled quantum field theory.

\bibliography{refs.bib}

\newpage

\section*{Methods}

The devices were fabricated using our ``pick-up method"~\cite{wang2013one} to achieve a multi-layer heterostructure with the TLG encapsulated by two flakes of hexagonal boron nitride (hBN) and thin graphite flakes as the top and bottom gates. The stacks were annealed under a high vacuum at 350 $^\circ$C for 2 hours. Electron-beam lithography was used to write a etch mask to define the Hall-bar geometry and the electrodes. Redundant regions were etched away by CHF$_{3}$/O$_{2}$ plasma~\cite{wang2013one}. Finally the TLG and gates were edge-contacted~\cite{wang2013one} by e-beam evaporating thin metal layers consisting of Cr/Pd/Au (1 nm/15 nm/100 nm). 
 
The transport measurements were performed in two systems, a dilution fridge with a base temperature of 15 mK and a VTI fridge down to 1.5 K, and both are with superconducting magnets. All data were taken using the standard four-terminal configuration with lock-in amplifier techniques by sourcing an AC current $I$ between 10 and 100 nA at a frequency of 17.777 Hz. The data of longitudinal conductance $\sigma_{xx}$ and Hall conductance $\sigma_{xy}$ are obtained from the measured resistances by $\sigma_{xx}=\rho_{xx}/(\rho_{xx}^2+R_{xy}^2)$ and $\sigma_{xy}=R_{xy}/(\rho_{xx}^2+R_{xy}^2)$.

\section*{Data availability}

The data that support the findings of this study are available from the corresponding authors upon request.

\section*{Code availability}

The code that support the findings of this study are available from the corresponding authors upon request.

\bigskip
\section*{Acknowledgments}

L.W. acknowledges support from the National Key Projects for Research and Development of China (Grant Nos. 2021YFA1400400, 2022YFA1204700), National Natural Science Foundation of China (Grant No. 12074173) and Natural Science Foundation of Jiangsu Province (Grant No. BK20220066). Y.H.W. acknowledges support from the National Natural Science Foundation of China (Grant No. 12174130). Z.Z. acknowledges the National Natural Science Foundation of China (Grant No.12074375), the Strategic Priority Research Program of CAS (Grant No.XDB33000000).  K.W. and T.T. acknowledge support from the JSPS KAKENHI (Grant Numbers 21H05233 and 23H02052) and World Premier International Research Center Initiative (WPI), MEXT, Japan. F.C.Z. acknowledges partial support from China Ministry of Sci and Tech (grant 2022YFA1403902), Priority Program of CAS grant No XDB28000000, NSFC grant No 11674278, and Chinese Academy of Sciences  under contract No. JZHKYPT-2021-08. G.L. acknowledges the support by the NSFC of China (grant No. 9206520), the National Basic Research Program of China from the MOST (grant No. 2022YFA1602803), and the Strategic Priority Research Program of the Chinese Academy of Sciences (grant No. XDB33010300). This work is supported by the Synergic Extreme Condition User Facility and by the Innovation Program for Quantum Science and Technology (Grant No. 2021ZD03026001).

\bigskip
\section*{Author Contributions}

L.W. conceived and designed the experiment. Y.C., Y.H., and Q.L. fabricated the samples. Y.C., Y.H., Q.L., B.T., G.L., and L.L. performed the transport measurements. Y.C., L.W., Y.H.W., F.C.Z., and Z.Z analyzed the data. Y.H.W. conducted theoretical analysis. K.W. and T.T supplied the hBN crystals. Y.C., Y.H.W. and L.W. wrote the manuscript with input from all co-authors.

\bigskip
\section*{Competing interests}

The authors declare no competing interest.


\newpage

\onecolumngrid

\setcounter{figure}{0}
\renewcommand{\figurename}{\textbf{Extended Data Fig.}}

\begin{figure*}[h!]
\begin{center}
\includegraphics[width=0.5\linewidth]{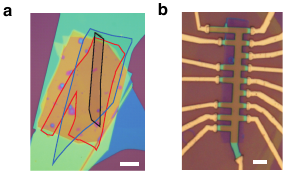}
\caption
{
\textbf{Optical image of one typical device.} 
       \textbf{a}, The optical image of the released stack after transferring. The blue (black) line denotes the boundary of top (bottom) graphite and the red line denotes the boundary of trilayer graphene (TLG). The scalebar corresponds to 10 $\mu$m.
       \textbf{b}, The completed device with Hall bar geometry that avoids the region with bubbles, after the deposition of Au/Pd/Cr contacts. The scalebar corresponds to 5 $\mu$m.
}
\label{fig:fig.e1}
\end{center}
\end{figure*}

\bigskip

\begin{figure*}[h!]
\includegraphics[width=0.6\linewidth]{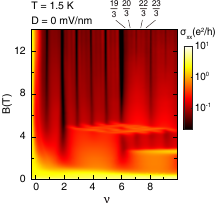}
\caption
{
\textbf{Quantum Hall states in the range $\boldsymbol{0{\leqslant}\nu{\leqslant}10}$}. 
      The color map of $\sigma_{xx}$ measured at at $D=0$ mV/nm and 1.5 K plotted versus $\nu$ and $B$. The $\nu=19/3, 23/3$ states appear at a lower $B$ compared to the integer quantum Hall state $\nu=7$, while the $\nu=20/3, 22/3$ states appear after it.
}
\label{fig:fig.e2}
\end{figure*}

\newpage

\begin{figure*}[h!]
\includegraphics[width=0.8\linewidth]{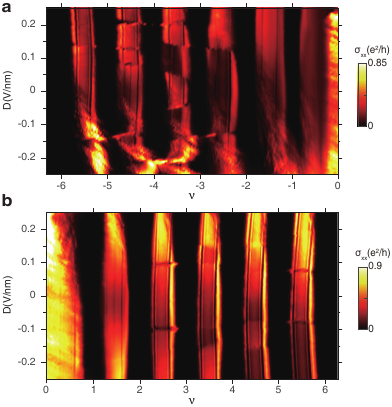}
\caption
{
\textbf{FQH states in the range $\boldsymbol{-6{\leqslant}\nu{\leqslant}6}$}. 
       \textbf{a}, The hole side. \textbf{b}, The electron side. Both panels are measured at $B=14$ T and 15 mK.
}
\label{fig:fig.e3}
\end{figure*}

\newpage

\begin{figure*}[h!]
\includegraphics[width=1\linewidth]{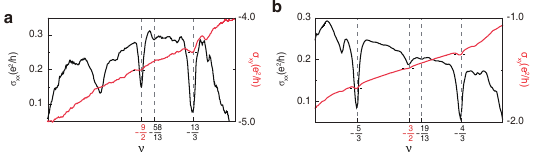}
\caption
{
\textbf{Evidence of the Levin-Halperin daughter states.} 
     \textbf{a-b}, $\sigma_{xx}$ and $\sigma_{xy}$ versus $\nu$ in the range $-5{\leqslant}\nu{\leqslant}-4$ ($B=14.9$ T and $D=0$ mV/nm in \textbf{a}) and $-2{\leqslant}\nu{\leqslant}-1$ ($B=14.6$ T and $D=217$ mV/nm in \textbf{b}). Both panels exhibit a small dip in $\sigma_{xx}$ at $\nu-[\nu]=7/13$ that may be interpreted as one of the Levin-Halperin daughter states.
}
\label{fig:fig.e4}
\end{figure*}

\begin{figure*}[h!]
\includegraphics[width=0.9\linewidth]{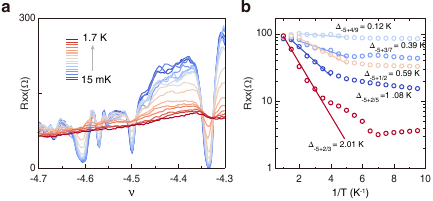}
\caption
{
\textbf{Arrhenius fitting for the gaps of FQH states in the range $\boldsymbol{-5{\leqslant}\nu{\leqslant}-4}$}.
       \textbf{a}, The linecuts of $R_{xx}$ versus $\nu$ at different temperatures $T$. 
       \textbf{b}, The Arrhenius fittings for $\nu=-5+2/5,3/7,4/9,1/2,2/3$ using the formula $R_{xx} \sim \exp (-\Delta/2T)$.
}
\label{fig:fig.e5}
\end{figure*}

\begin{figure*}[h!]
\includegraphics[width=1\linewidth]{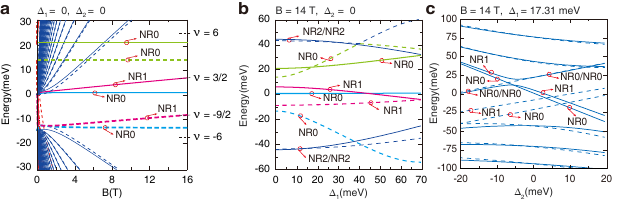}
\caption
{
\textbf{LL diagrams obtained by numerical calculations.}
       \textbf{a}, The same diagram as in Fig.~3f of the main text but plotted in a different way here for comparison. The solid (dashed) lines represent the $\mathbf{K}_{+}$ ($\mathbf{K}_{-}$) valley. The orbital content of a level is indicated as NR$x$ if it has a substantial weight in the non-relativistic LL with index $x$. 
       \textbf{b}, The evolution of LLs with $\Delta_{1}$ when $B=14$ T and $\Delta_{2}$=0. The MLG and BLG levels in panel \textbf{a} are hybridized by $\Delta_{1}$. The green $\mathbf{K}_{\pm},N_{\rm M}=0$ levels cross with each other at $\Delta_{1} \approx 17$ meV. The blue $\mathbf{K}_{+},N_{\rm B}=0$ level and the red $\mathbf{K}_{+},N_{\rm B}=1$ level cross with each other at $\Delta_{1} \approx 40$ meV. The blue $\mathbf{K}_{-},N_{\rm B}=0$ level crosses with two NR2 levels at $\Delta_{1} \approx 40$ meV. 
       \textbf{c}, The evolution of LLs with $\Delta_{2}$ when $B=14$ T and $\Delta_{1}=17.31$ meV. One NR1 level at $\Delta_{2}=-10$ meV that may host the $\nu=9/2$ state is indicated. One NR1 level in the $\mathbf{K}_{-}$ valley crosses with one NR0 level in the $\mathbf{K}_{+}$ valley at $\Delta_{2} \approx 3.5$ meV. More details about these calculations are given in the Supplementary Information.
}
\label{fig:fig.e6}
\end{figure*}


\renewcommand{\figurename}{\textbf{Fig.}}

\clearpage
\centerline{\Large{Supplementary Information for}}
\vspace*{1\baselineskip} 

\centerline{\textbf{\large{Tunable even- and odd-denominator fractional quantum Hall states }}}
\centerline{\textbf{\large{in trilayer graphene}}}
\vspace*{1\baselineskip} 

\centerline{Yiwei Chen \textit{et al.}}
\vspace*{1\baselineskip} 

\setcounter{figure}{0}
\setcounter{equation}{0}
\renewcommand\thefigure{S\arabic{figure}}
\renewcommand\theequation{S\arabic{equation}}

This Supplementary Information provides more details about the theoretical analysis of fractional quantum Hall states in trilayer graphene.

\section{Models}

To begin with, we define the single-particle tight-binding model for trilayer graphene (TLG). There are two sublattice sites $A$ and $B$ in each layer of TLG and their layer indices are appended as subscripts $1,2,3$. The distance between Carbon atoms is $a=0.142$ nm and the lattice constant is $\widetilde{a}=\sqrt{3}a$. As illustrated in Fig.~1 of the main text, six hopping constants $\gamma_{i}$ ($i=0,1,\cdots,5$) are considered in the Slonczewski-Weiss-McClure (SWMc) parametrization. The distance between layers along the vertical direction is $d=0.335$ nm. An additional parameter $\delta$ is used to account for the onsite potential of $B_{1}$, $A_{2}$, and $B_{3}$ lattice sites. If a displacement field is present, the potential difference between the top and bottom layer is quantified by $\Delta_{1}$. Based on symmetry consideration, it was proposed another parameter $\Delta_{2}$ should be introduced to describe the deviation of potential on the middle layer from the mean of the potentials on top and bottom layers. In principle, $\Delta_{2}$ could be nonzero even when $\Delta_{1}$ is zero. Let us choose the basis to be $A_{1},B_{1},A_{2},B_{2},A_{3},B_{3}$. The tight-binding Hamiltonian is 
\begin{eqnarray}
\mathcal{H} = \left[ \begin{matrix}
\Delta_{1}+\Delta_{2} & \gamma_{0} t^{*}(\mathbf{k}) & \gamma_{4} t^{*}(\mathbf{k}) & \gamma_{3} t(\mathbf{k}) & \gamma_{2}/2 & 0 \\
\gamma_{0} t(\mathbf{k}) & \delta+\Delta_{1}+\Delta_{2} & \gamma_{1} & \gamma_{4} t^{*}(\mathbf{k}) & 0 & \gamma_{5}/2 \\
\gamma_{4} t(\mathbf{k}) & \gamma_{1} & \delta-2\Delta_{2} & \gamma_{0} t^{*}(\mathbf{k}) & \gamma_{4} t(\mathbf{k}) & \gamma_{1} \\
\gamma_{3} t^{*}(\mathbf{k}) & \gamma_{4} t(\mathbf{k}) & \gamma_{0} t(\mathbf{k}) & -2\Delta_{2} & \gamma_{3}t^{*}(\mathbf{k}) & \gamma_{4} t(\mathbf{k}) \\
\gamma_{2}/2 & 0 & \gamma_{4} t^{*}(\mathbf{k}) & \gamma_{3} t(\mathbf{k}) & -\Delta_{1}+\Delta_{2} & \gamma_{0} t^{*}(\mathbf{k}) \\
0 & \gamma_{5}/2 & \gamma_{1} & \gamma_{4} t^{*}(\mathbf{k}) & \gamma_{0} t(\mathbf{k}) & \delta-\Delta_{1}+\Delta_{2}
\end{matrix} \right]
\end{eqnarray}
where $t(\mathbf{k}) = -1-2\cos(k_{x}\widetilde{a}/2)\exp(i\sqrt{3}k_{y}\widetilde{a}/2)$ is a summation over nearest neighbors. We adopt the parameters $\gamma_{0}=3.1$, $\gamma_{1}=0.39$, $\gamma_{2}=-0.028$, $\gamma_{3}=0.315$, $\gamma_{4}=0.041$, $\gamma_{5}=0.05$, and $\delta=0.046$ (all in units of eV) for our calculations~\cite{serbyn2013new}. The band structure with $\Delta_{1}=\Delta_{2}=0$ is presented in Fig.~1 of the main text.

Next we turn to the Landau levels (LLs) of TLG. The energy bands have two valleys in the hexagonal Brillouin zone with momentum $\mathbf{K}_{\pm}=({\pm}4\pi/3,0)$. In the vicinity of these valleys, the Hamiltonian can be expanded to yield
\begin{eqnarray}
\mathcal{H}_{\mathbf{K}_{+}} = && \left[ \begin{matrix}
\Delta_{1}+\Delta_{2} & v_{0}\pi^{-} & v_{4}\pi^{-} & v_{3}\pi^{+} & \gamma_{2}/2 & 0 \\
v_{0}\pi^{+} & \delta+\Delta_{1}+\Delta_{2} & \gamma_{1} & v_{4}\pi^{-} & 0 & \gamma_{5}/2 \\
v_{4}\pi^{+} & \gamma_{1} & \delta-2\Delta_{2} & v_{0}\pi^{-} & v_{4}\pi^{+} & \gamma_{1} \\
v_{3}\pi^{-} & v_{4}\pi^{+} & v_{0}\pi^{+} & -2\Delta_{2} & v_{3}\pi^{-} & v_{4}\pi^{+} \\
\gamma_{2}/2 & 0 & v_{4}\pi^{-} & v_{3}\pi^{+} & -\Delta_{1}+\Delta_{2} & v_{0}\pi^{-} \\
0 & \gamma_{5}/2 & \gamma_{1} & v_{4}\pi^{-} & v_{0}\pi^{+} & \delta-\Delta_{1}+\Delta_{2}
\end{matrix} \right]
\end{eqnarray}
and
\begin{eqnarray}
\mathcal{H}_{\mathbf{K}_{-}} = && \left[ \begin{matrix}
\Delta_{1}+\Delta_{2} & -v_{0}\pi^{+} & -v_{4}\pi^{+} & -v_{3}\pi^{-} & \gamma_{2}/2 & 0 \\
-v_{0}\pi^{-} & \delta+\Delta_{1}+\Delta_{2} & \gamma_{1} & -v_{4}\pi^{+} & 0 & \gamma_{5}/2 \\
-v_{4}\pi^{-} & \gamma_{1} & \delta-2\Delta_{2} & -v_{0}\pi^{+} & -v_{4}\pi^{-} & \gamma_{1} \\
-v_{3}\pi^{+} & -v_{4}\pi^{-} & -v_{0}\pi^{-} & -2\Delta_{2} & -v_{3}\pi^{+} & -v_{4}\pi^{-} \\
\gamma_{2}/2 & 0 & -v_{4}\pi^{+} & -v_{3}\pi^{-} & -\Delta_{1}+\Delta_{2} & -v_{0}\pi^{+} \\
0 & \gamma_{5}/2 & \gamma_{1} & -v_{4}\pi^{+} & -v_{0}\pi^{-} & \delta-\Delta_{1}+\Delta_{2}
\end{matrix} \right]
\end{eqnarray}
where ${\hbar}v_{i}=3a\gamma_{i}/2$, $\pi^{-}=\hbar(k_{x}-ik_{y})$, and $\pi^{+}=\hbar(k_{x}+ik_{y})$. For our purpose, it is convenient to change the basis to 
\begin{eqnarray}
\frac{A_{1}-A_{3}}{\sqrt{2}} , \frac{B_{1}-B_{3}}{\sqrt{2}}, \frac{A_{1}+A_{3}}{\sqrt{2}}, \frac{B_{1}+B_{3}}{\sqrt{2}}, A_{2}, B_{2}
\end{eqnarray}
such that $\mathcal{H}_{\mathbf{K}_{\pm}}$ has the block form
\begin{eqnarray}
\left[ \begin{matrix}
\mathcal{H}_{\rm MLG} & \mathcal{H}_{\rm mix} \\
\mathcal{H}^{\dag}_{\rm mix} & \mathcal{H}_{\rm BLG} \\
\end{matrix} \right]
\end{eqnarray}
with a two-dimensional monolayer graphene (MLG) part $\mathcal{H}_{\rm MLG}$, a four-dimensional bilayer graphene (BLG) part $\mathcal{H}_{\rm BLG}$, and
a mixing part 
\begin{eqnarray}
\mathcal{H}_{\rm mix} = \left[ \begin{matrix}
\Delta_{1} & 0 & 0 & 0 \\
0 & \Delta_{1}  & 0 & 0 \\
\end{matrix} \right]
\end{eqnarray}
This representation clearly shows that the TLG Hamiltonian can be decomposed to a combination of MLG and BLG parts in the absence of $\Delta_{1}$. In the $\mathbf{K}_{+}$ valley, we have
\begin{eqnarray}
\mathcal{H}^{+}_{\rm MLG} = \left[ \begin{matrix}
-\frac{\gamma_{2}}{2}+\Delta_{2} & v_{0}\pi^{-} \\
v_{0}\pi^{+} & -\frac{\gamma_{5}}{2}+\delta+\Delta_{2}
\end{matrix} \right] \qquad
\mathcal{H}^{+}_{\rm BLG} = \left[ \begin{matrix}
\frac{\gamma_{2}}{2}+\Delta_{2} & v_{0}\pi^{-} & -\sqrt{2}v_{4}\pi^{-} & \sqrt{2}v_{3}\pi^{+} \\
v_{0}\pi^{+} & \frac{\gamma_{5}}{2}+\delta+\Delta_{2} & \sqrt{2}\gamma_{1} & -\sqrt{2}v_{4}\pi^{-} \\
-\sqrt{2}v_{4}\pi^{+} & \sqrt{2}\gamma_{1} & \delta-2\Delta_{2} & v_{0}\pi^{-} \\
\sqrt{2}v_{3}\pi^{-} & -\sqrt{2}v_{4}\pi^{+} & v_{0}\pi^{+} & -2\Delta_{2}
\end{matrix} \right].
\end{eqnarray}
In the $\mathbf{K}_{-}$ valley, we have
\begin{eqnarray}
\mathcal{H}^{-}_{\rm MLG} = \left[ \begin{matrix}
-\frac{\gamma_{2}}{2}+\Delta_{2} & -v_{0}\pi^{+} \\
-v_{0}\pi^{-} & -\frac{\gamma_{5}}{2}+\delta+\Delta_{2}
\end{matrix} \right] \qquad 
\mathcal{H}^{-}_{\rm BLG} = \left[ \begin{matrix}
\frac{\gamma_{2}}{2}+\Delta_{2} & -v_{0}\pi^{+} & \sqrt{2}v_{4}\pi^{+} & -\sqrt{2}v_{3}\pi^{-} \\
-v_{0}\pi^{-} & \frac{\gamma_{5}}{2}+\delta+\Delta_{2} & \sqrt{2}\gamma_{1} & \sqrt{2}v_{4}\pi^{+} \\
\sqrt{2}v_{4}\pi^{-} & \sqrt{2}\gamma_{1} & \delta-2\Delta_{2} & -v_{0}\pi^{+} \\
-\sqrt{2}v_{3}\pi^{+} & \sqrt{2}v_{4}\pi^{-} & -v_{0}\pi^{-} & -2\Delta_{2}
\end{matrix} \right].
\end{eqnarray}
If we turn on a perpendicular magnetic field generated by the vector potential $\mathbf{A}$, miniaml coupling is achieved by the substitution
\begin{eqnarray}
\pi^{-} \rightarrow \Pi^{-}=\pi^{-}-e(A_{x}-iA_{y}), \quad \pi^{+} \rightarrow \Pi^{+}=\pi^{+}-e(A_{x}+iA_{y}).
\end{eqnarray}
In the non-relativistic Landau problem, interlevel ladder operators are defined as
\begin{eqnarray}
\widehat{\mathsf{a}}=\frac{\ell_{B}}{\sqrt{2}\hbar} \Pi^{-} \qquad \widehat{\mathsf{a}}^{\dag}=\frac{\ell_{B}}{\sqrt{2}\hbar} \Pi^{+}
\end{eqnarray}
where $\ell_{B}=\sqrt{\hbar/(eB)}$ is the magnetic length. For our TLG Hamiltonian $\mathcal{H}_{\mathbf{K}^{\pm}}$, this amounts to the replacement
\begin{eqnarray}
\gamma_{i}\Pi^{-} \rightarrow \frac{3\gamma_{i}a}{\sqrt{2}\ell_{B}} \mathsf{a}, \quad \gamma_{i}\Pi^{+} \rightarrow \frac{3\gamma_{i}a}{\sqrt{2}\ell_{B}} \mathsf{a}^{\dag}.
\end{eqnarray}
The number operator $\widehat{\mathsf{a}}^{\dag}\widehat{\mathsf{a}}$ can be defined as usual, and its eigenstates $|n\rangle$ satisfy
\begin{eqnarray}
\widehat{\mathsf{a}} |n\rangle = \sqrt{n} |n-1\rangle, \quad \widehat{\mathsf{a}}^{\dag} |n\rangle = \sqrt{n+1} |n+1\rangle
\end{eqnarray}
The eigenstates of $\mathcal{H}_{\mathbf{K}_{\pm}}$ can be expressed as
\begin{eqnarray}
\begin{bmatrix}
f_{00} |0\rangle + f_{01} |1\rangle + f_{02} |2\rangle + \ldots \\
f_{10} |0\rangle + f_{11} |1\rangle + f_{12} |2\rangle + \ldots \\
f_{20} |0\rangle + f_{21} |1\rangle + f_{22} |2\rangle + \ldots \\
f_{30} |0\rangle + f_{31} |1\rangle + f_{32} |2\rangle + \ldots \\
f_{40} |0\rangle + f_{41} |1\rangle + f_{42} |2\rangle + \ldots \\
f_{50} |0\rangle + f_{51} |1\rangle + f_{52} |2\rangle + \ldots \\
\end{bmatrix},
\label{eq:LandauEigenState}
\end{eqnarray}
where each row of this spinor is an infinite summation and cannot be computed exactly. In our calculation, a proper truncation is imposed such that the summation terminates at a finite $n$. If we are only interested in the eigenvalues close to zero, a moderate $n$ would be sufficient. For example, the Landau level diagrams are obtained for $n=19$. On the other hand, the many-body problem uses $n=7$.

It should be emphasized that some aspects of the energy spectrum depend sensitively on the system parameters. A precise determination of the parameters is difficult, but a certain degree of consensus has emerged in the past decade. In the absence of $\Delta_{1}$, the LLs can be labeled as MLG and BLG ones as we have done in Fig.~3f of the main text. More generally, when the MLG and BLG parts are hybridized by nonzero $\Delta_{1}$, it is more useful to inspect the orbital content of each level. If $|0\rangle$ has the largest weight in Eq.~\ref{eq:LandauEigenState}, it would be denoted as non-relativistic 0 (NR0). If $|1\rangle$ has a substantial weight in Eq.~\ref{eq:LandauEigenState}, it would be denoted as non-relativistic 1 (NR1). Naturally, the $\mathbf{K}_{\pm},N_{\rm B}=0$ levels are of the NR0 type and the $\mathbf{K}_{\pm},N_{\rm B}=1$ levels are of the NR1 type. The effects of changing the parameters shall be discussed when we study the many-body problem. The Zeemann coupling is $E_{\rm Z} = g \mu_{B} B = 0.058g \; B[\text{Tesla}] \; \text{meV}$ with $g=2$. It is $1.62$ meV at $B=14$ T and is smaller than the energy separation between the NR0 and NR1 levels. The hexagonal boron-nitride has dielectric constants $\varepsilon^{\perp}_{\rm BN}=3$ in the perpendicular direction and $\varepsilon^{\parallel}_{\rm BN}=6.6$ within the two-dimensional plane. The interaction between electrons is the screened Coulomb potential 
\begin{eqnarray}
V_{\rm SC}(\mathbf{q}) = \frac{e^{2}}{4\pi\varepsilon_{0}\varepsilon^{\parallel}_{\rm BN}\ell_{B}} \frac{2\pi}{q} \tanh(qd) = \frac{2\pi}{q} \tanh(qd) \; \frac{56.2}{\varepsilon^{\parallel}_{\rm BN}} \; \sqrt{B[\text{Tesla}]} \; \text{meV},
\end{eqnarray}
when $d$ is the distance between top and bottom graphite gates.

\section{Methods}

In our numerical calculation, electrons are placed on a rectangular torus whose two sides are described by the vectors 
\begin{eqnarray}
\mathbf{L}_{1} = L_{1}\widehat{e}_{v} \qquad \mathbf{L}_{2} = L_{2}\widehat{e}_{y}
\end{eqnarray}
The flux magnetic flux $N_{\phi}$ through the torus is quantized to satsify $L_{1}L_{2} = 2\pi \ell^{2}_{B} N_{\phi}$. Let us consider non-relativistic electrons with anisotropic mass as described by the single-particle Hamiltonian
\begin{eqnarray}
\mathcal{H}_{\rm NR}(\Lambda) = \frac{1}{2M} \left[ \Lambda p^{2}_{x} + \frac{1}{\Lambda} \left( p_{y} - eBx \right)^{2} \right].
\end{eqnarray}
The single-particle wave functions in the LLs are
\begin{eqnarray}
\phi^{\alpha}_{m}(\mathbf{r}) &=& \frac{1}{ \left( 2^{\alpha}\alpha! \sqrt{\pi}\ell_{B}L_{2} \sqrt{\Lambda} \right)^{1/2}} \sum^{\mathbb{Z}}_{k} \exp \left\{ -\frac{1}{2\Lambda} \left[ \frac{x}{\ell_{B}} - \frac{2{\pi}\ell_{B}}{L_{2}} \left( m+kN_{\phi} \right) \right]^{2} + i \frac{2{\pi}y}{L_{2}} \left( m+kN_{\phi} \right) \right\} \nonumber \\
&\times& H_{\alpha} \left\{ \frac{1}{\sqrt{\Lambda}} \left[ \frac{x}{\ell_{B}} - \frac{2{\pi}\ell_{B}}{L_{2}} \left( m+kN_{\phi} \right) \right] \right\},
\end{eqnarray}
where $\alpha\in\mathbb{N}$ is the Landau level index and $m\in[0,\ldots,N_{\phi}-1]$. For a generic two-body interaction potential $V(\mathbf{r}_{1}-\mathbf{r}_{2})$, the many-body Hamiltonian can be written as
\begin{eqnarray}
H_{\rm MB}(\Lambda) = \frac{1}{2L_{1}L_{2}} \sum_{\mathbf{q}} V(\mathbf{q}) : \rho(\mathbf{q}) \rho(-\mathbf{q}) :
\end{eqnarray}
with
\begin{eqnarray}
V(\mathbf{r}_{1}-\mathbf{r}_{2}) = \frac{1}{L_{1}L_{2}} \sum_{\mathbf{q}} V(\mathbf{q}) \; \exp \left[ -i\mathbf{q} \cdot (\mathbf{r}_{1} - \mathbf{r}_{2}) \right] \qquad \mathbf{q} = \frac{2\pi}{L_{1}} q_{1} \widehat{e}_{x} + \frac{2\pi}{L_{2}} q_{2} \widehat{e}_{y} 
\end{eqnarray}
Based on previous experience, it is useful to define 
\begin{eqnarray}
|\mathbf{q}(\Lambda)|^{2} = \Lambda \left( \frac{2{\pi}\ell_{B}}{L_{1}\sin\theta}q_{1} - \frac{2{\pi}\ell_{B}\cot\theta}{L_{2}}q_{2} \right)^{2} + \frac{1}{\Lambda} \left( \frac{2{\pi}\ell_{B}}{L_{2}}q_{2} \right)^{2}
\end{eqnarray}
and the form factor
\begin{eqnarray}
&& F_{\alpha\beta}(q_{1},q_{2},\Lambda) = \frac{1}{\sqrt{\alpha!\beta!2^{\alpha+\beta}}} \sum^{{\rm min}[\alpha,\beta]}_{k} 2^{k}k! \; \binom{\alpha}{k} \; \binom{\beta}{k} \left[ i \sqrt{\Lambda} \frac{2{\pi}\ell_{B}}{L_{1}}q_{1} - \frac{2\pi\ell_{B}}{\sqrt{\Lambda}L_{2}} q_{2} \right]^{\alpha-k} \left[ i \sqrt{\Lambda} \frac{2{\pi}\ell_{B}}{L_{1}}q_{1} + \frac{2\pi\ell_{B}}{\sqrt{\Lambda}L_{2}} q_{2} \right]^{\beta-k}
\end{eqnarray}
In subsequent calculations, we will need their derivatives
\begin{eqnarray}
\left. \frac{d|\mathbf{q}(\Lambda)|^{2}}{d\Lambda} \right|_{\Lambda=1} = \left( \frac{2{\pi}\ell_{B}}{L_{1}}q_{1} \right)^{2} - \left( \frac{2{\pi}\ell_{B}}{L_{2}}q_{2} \right)^{2}
\label{eq:GeneralDerive1}
\end{eqnarray}
and
\begin{eqnarray}
&& \left. \frac{dF_{\alpha\beta}(q_{1},q_{2})}{d\Lambda} \right|_{\Lambda=1} = \frac{1}{\sqrt{\alpha!\beta!2^{\alpha+\beta}}} \sum^{{\rm min}[\alpha,\beta]}_{k} 2^{k}k! \; \binom{\alpha}{k} \; \binom{\beta}{k} \nonumber \\
&\times& \left\{ \frac{1}{2} (\alpha-k) \left( i \frac{2{\pi}\ell_{B}}{L_{1}}q_{1} - \frac{2\pi\ell_{B}}{L_{2}} q_{2} \right)^{\alpha-k-1} \left( i \frac{2{\pi}\ell_{B}}{L_{1}}q_{1} + \frac{2\pi\ell_{B}}{L_{2}} q_{2} \right)^{\beta-k+1} \right. \nonumber \\
&+& \left. \frac{1}{2} (\beta-k) \left( i \frac{2{\pi}\ell_{B}}{L_{1}}q_{1} - \frac{2\pi\ell_{B}}{L_{2}} q_{2} \right)^{\alpha-k+1} \left( i \frac{2{\pi}\ell_{B}}{L_{1}}q_{1} + \frac{2\pi\ell_{B}}{L_{2}} q_{2} \right)^{\beta-k-1} \right\}
\label{eq:GeneralDerive2}
\end{eqnarray} 
If $\alpha=\beta$, the second derivative can be simplified to 
\begin{eqnarray}
&& \left. \frac{dF_{\alpha\alpha}(q_{1},q_{2})}{d\Lambda} \right|_{\Lambda=1} = - \left[ \left( \frac{2{\pi}\ell_{B}}{L_{1}}q_{1} \right)^{2} - \left( \frac{2\pi\ell_{B}}{L_{2}} q_{2} \right)^{2} \right] \nonumber \\
&\times& \frac{1}{\sqrt{\alpha!\alpha!2^{2\alpha}}} \sum^{\alpha-1}_{k} 2^{k}k! \; \binom{\alpha}{k} \; \binom{\alpha}{k} (\alpha-k) \left[ i \frac{2{\pi}\ell_{B}}{L_{1}}q_{1} - \frac{2\pi\ell_{B}}{L_{2}} q_{2} \right]^{\alpha-k-1} \left[ i \frac{2{\pi}\ell_{B}}{L_{1}}q_{1} + \frac{2\pi\ell_{B}}{L_{2}} q_{2} \right]^{\alpha-k-1}.
\label{eq:SpecialDerive}
\end{eqnarray} 

To study the FQH states, we keep two LLs with single-particle energy $\epsilon^{(\alpha)}$ and single-particle wave functions
\begin{eqnarray}
|\widetilde{\phi}^{(\alpha)}\rangle = \begin{bmatrix}
\sum^{7}_{n=0} f^{(\alpha)}_{0n} |n\rangle \\
\sum^{7}_{n=0} f^{(\alpha)}_{1n} |n\rangle \\
\sum^{7}_{n=0} f^{(\alpha)}_{2n} |n\rangle \\
\sum^{7}_{n=0} f^{(\alpha)}_{3n} |n\rangle \\
\sum^{7}_{n=0} f^{(\alpha)}_{4n} |n\rangle \\
\sum^{7}_{n=0} f^{(\alpha)}_{5n} |n\rangle \\
\end{bmatrix}.
\end{eqnarray} 
The superscript $\alpha=0,1$ labels the two levels, the first subscript $0,1,\ldots,5$ labels the components of the spinor, and the second subscript $n=0,1,\cdots,7$ denotes the non-relativistic LL indices. These two levels are called active and other levels are neglected. The creation (annihilation) operator associated with $|\widetilde{\phi}^{(\alpha)}\rangle$ is denoted as $C^{\dag}_{{\alpha}m}$ ($C_{{\alpha}m}$). The many-body Hamiltonian is
\begin{eqnarray}
H_{\rm MB}(\Lambda) &=& \sum_{\alpha,m} \epsilon^{(\alpha)} C^{\dag}_{{\alpha}m} C_{{\alpha}m} + \frac{1}{2L_{1}L_{2}} \sum_{\{\alpha_{i}\}} \sum_{\{m_{i}\}} \sum_{q_{1},q_{2}} V_{\rm SC}(\mathbf{q}) \exp \left[ -\frac{1}{2} |\mathbf{q}(\Lambda)|^{2} \ell^{2}_{B} - i\frac{2{\pi}q_{1}}{N_{\phi}} \left( m_{1}-m_{4} \right) \right] \nonumber \\
&\times& \widetilde{F}_{\alpha_{1}\alpha_{3}}(-q_{1},-q_{2},\Lambda) \widetilde{F}_{\alpha_{2}\alpha_{4}}(q_{1},q_{2},\Lambda) \; \widetilde{\delta}_{m_{1},m_{3}-q_{2}} \widetilde{\delta}_{m_{2},m_{4}+q_{2}} \; C^{\dag}_{\alpha_{1}m_{1}} C^{\dag}_{\alpha_{2}m_{2}} C_{\alpha_{4}m_{4}} C_{\alpha_{3}m_{3}}
\end{eqnarray}
where $\widetilde{\delta}$ is a generalized Kroneker function defined as
\begin{eqnarray}
\widetilde{\delta}_{s,t+q_{2}} = 1 \qquad \text{if and only if} \qquad s \; \text{mod} \; N_{\phi} = (t+q_{2}) \; \text{mod} \; N_{\phi}
\end{eqnarray}
and
\begin{eqnarray}
\widetilde{F}_{\alpha_{1}\alpha_{3}}(-q_{1},-q_{2},\Lambda) &=& \sum^{5}_{i=0} \sum^{7}_{n_{1},n_{2}} f^{(\alpha_{1})*}_{in_{1}} f^{(\alpha_{3})}_{in_{2}} F_{n_{1}n_{2}}(-q_{1},-q_{2},\Lambda) \nonumber \\
\widetilde{F}_{\alpha_{2}\alpha_{4}}(q_{1},q_{2},\Lambda) &=& \sum^{5}_{i=0} \sum^{7}_{n_{1},n_{2}} f^{(\alpha_{2})*}_{in_{1}} f^{(\alpha_{4})}_{in_{2}} F_{n_{1}n_{2}}(q_{1},q_{2},\Lambda).
\end{eqnarray}
It is obvious that the total momentum $Y=\sum_{i} m_{i}$ is conserved. The low-energy eigenstates of $\widehat{H}_{\rm MB}(1)$ are computed by exact diagonalization. Due to the exponential growth of Hilbert space dimension, the maximal number of electrons is severely constrained. This problem can be partially mitigated if we impose a constraint on the number of electrons in the higher LL. If the total filling factor of the two levels is $3/2$ and there is no interaction, then the lower level is fully occupied and the higher level is half filled. As interaction is turned on, it is possible for all electrons to move to the higher level, but this configuration is not energetically favored. In our calculation, we only retain the configurations in which the higher level does not deviate too much from half filling.

The purpose of studying general $\Lambda$ is to define chiral graviton operators and compute their spectral functions~\cite{liou2019}. If we choose $\Lambda=1+\xi$ and treat $\xi$ as a small parameter, the Hamiltonian can be expanded as
\begin{eqnarray}
H_{\rm MB}(\Lambda) = H_{\rm MB}(1) + \xi \mathcal{O}
\end{eqnarray}
with graviton operator
\begin{eqnarray}
\mathcal{O} &=& \frac{1}{2L_{1}L_{2}} \sum_{\{\alpha_{i}\}} \sum_{\{m_{i}\}} \sum_{q_{1},q_{2}} \; V(\mathbf{q}) \exp \left[ - \frac{1}{2} |\mathbf{q}|^{2}\ell^{2}_{B} - i\frac{2{\pi}q_{1}}{N_{\phi}} (m_{1}-m_{4}) \right] \nonumber \\
&\times& \widetilde{\delta}_{m_{1},m_{3}-q_{2}} \widetilde{\delta}_{m_{2},m_{4}+q_{2}} \; C^{\dag}_{\alpha_{1}m_{1}} C^{\dag}_{\alpha_{2}m_{2}} C_{\alpha_{4}m_{4}} C_{\alpha_{3}m_{3}} \nonumber \\
&\times& \left\{ -\frac{1}{2} \left[ \left( \frac{2{\pi}\ell_{B}}{L_{1}}q_{1} \right)^{2} - \left( \frac{2{\pi}\ell_{B}}{L_{2}}q_{2} \right)^{2} \right] \widetilde{F}_{\alpha_{1}\alpha_{3}}(-q_{1},-q_{2},1) \widetilde{F}_{\alpha_{2}\alpha_{4}}(q_{1},q_{2},1) \right. \nonumber \\
&\phantom{=}& \left. + \left. \frac{d\widetilde{F}_{\alpha_{1}\alpha_{3}}(-q_{1},-q_{2},\Lambda)}{d\Lambda} \right|_{\Lambda=1} \widetilde{F}_{\alpha_{2}\alpha_{4}}(q_{1},q_{2},1) + \widetilde{F}_{\alpha_{1}\alpha_{3}}(-q_{1},-q_{2},1) \left. \frac{d\widetilde{F}_{\alpha_{2}\alpha_{4}}(q_{1},q_{2},\Lambda)}{d\Lambda} \right|_{\Lambda=1} \right\}.
\end{eqnarray}
It is more complicated than previously used ones because we have two LLs here~\cite{liou2019,haldane2021}. While the physics of gravitons in two LLs may be very interesting, it is more convenient to simplify the problem by going back to one level. If the total filling factor of the two LLs is $3/2$, the eigenstates obtained by exact diagonalization will be projected to the higher level. The summation over $\alpha$ in $\mathcal{O}$ will be limited to the higher level. The operator thus obtained is still not good enough because it cannot reveal the graviton chirality. To this end, we change the prefactors in Eqs.~\eqref{eq:GeneralDerive1} and~\eqref{eq:SpecialDerive} as
\begin{eqnarray}
\left[ \left( \frac{2{\pi}\ell_{B}}{L_{1}}q_{1} \right)^{2} - \left( \frac{2\pi\ell_{B}}{L_{2}} q_{2} \right)^{2} \right] \quad \Rightarrow \quad \left( \frac{2{\pi}\ell_{B}}{L_{1}}q_{1} \mp \frac{2\pi\ell_{B}}{L_{2}} q_{2} \right)^{2}
\end{eqnarray}
to define chiral graviton operators $\mathcal{O}_{\pm}$. This replacement is motivated by the connection between chiral graviton operators and anisotropic Haldane pseudopotentials~\cite{yangbo2017}. 

The eigenstates of $\widehat{H}_{\rm MB}(1)$ are denoted as $|\Psi_{n}\rangle$ and the associated eigenvalues are $E_{n}$. The total weights of the chiral graviton operators are
\begin{eqnarray}
W_{\pm} = \langle \Psi_{0} | \mathcal{O}^{\dag}_{\pm} \mathcal{O}_{\pm} | \Psi_{0} \rangle
\end{eqnarray}
and the normalized spectral functions are
\begin{eqnarray}
I_{\pm}(\omega) = \sum_{n} \frac{\left| \langle \Psi_{n} | \mathcal{O}_{\pm} | \Psi_{0} \rangle \right|^{2}}{W_{\pm}} \delta(\omega - E_{n} + E_{0}).
\end{eqnarray}
In finite-size systems with discrete energy levels, $I_{\pm}(\omega)$ can be computed using a method based on Lanczos tridiagonalizatio and continued fraction~\cite{gagliano1987}. We introduce a rescaled operator $\mathcal{F}=\mathcal{O}_{\pm}/\sqrt{W_{\pm}}$ and define its Green's function
\begin{eqnarray}
G(t) = \langle \Psi_{0} | \exp(iH_{\rm MB}t) \mathcal{F}^{\dag} \exp(-iH_{\rm MB}t) \mathcal{F} | \Psi_{0} \rangle.
\end{eqnarray}
Fourier transform yields
\begin{eqnarray}
G(\omega) = \int dt \; \exp(i{\omega}t) G(t) = \sum_{n} \frac{\left| \langle \Psi_{n} | \mathcal{F} | \Psi_{0} \rangle \right|^{2}}{\omega - E_{n} + E_{0}} = \sum_{n} \frac{w^{2}_{n}}{\omega - p_{n}}
\end{eqnarray}
where $p_{n}$ are the poles and $w^{2}_{n}$ are the weights. Using $|F_{0}\rangle=\mathcal{F}|\Psi_{0}\rangle$ as the initial vector, the Lanczos algorithm generates a tridiagonal matrix. $E_{0}+p_{n}$ is the $n$-th eigenvalue of this matrix and $\omega_{n}$ is the first element of the $n$-th eigenstate.

\section{Results}

\begin{figure}[ht]
\includegraphics[width=0.80\textwidth]{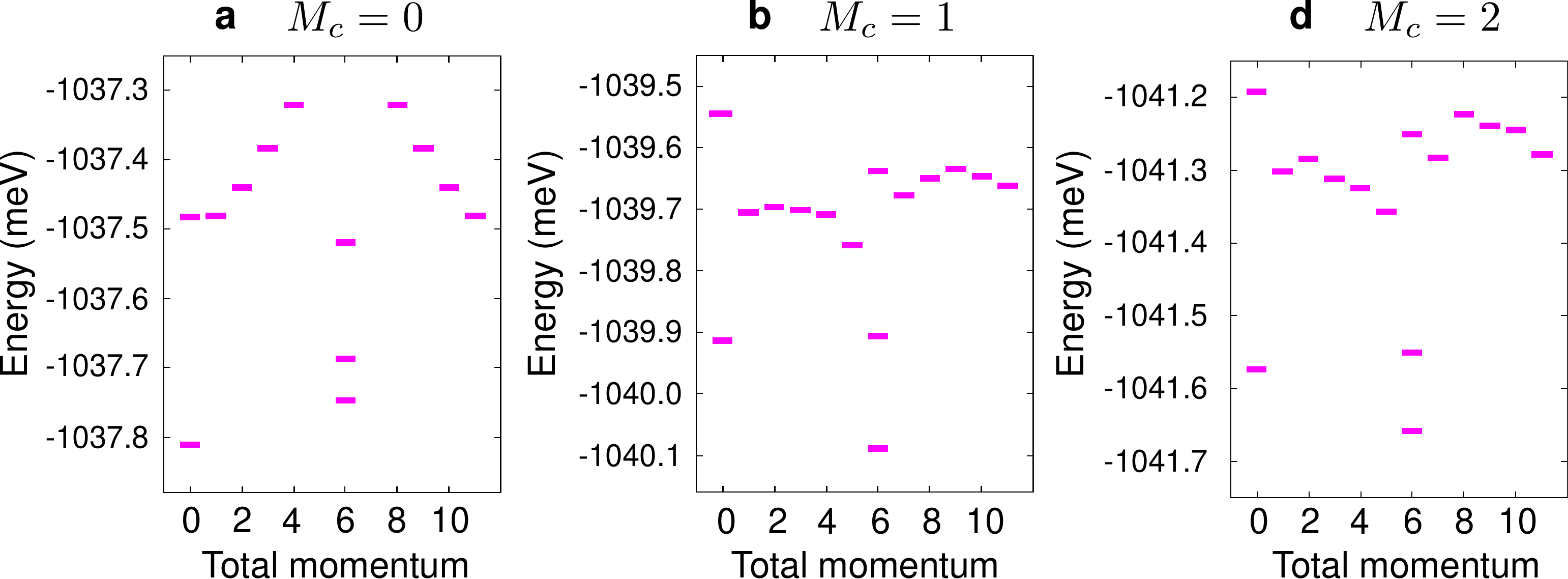}
\caption{\textbf{Low-energy eigenvalues of the $\boldsymbol{\nu=-9/2}$ state.} There are 36 electrons in total and the number of electrons that is allowed to escape from $\mathbf{K}_{-},N_{\rm B}=0$ is denoted as $M_{c}$.}
\label{Fig:Theory1}
\end{figure}

\begin{figure}[ht]
\includegraphics[width=0.95\textwidth]{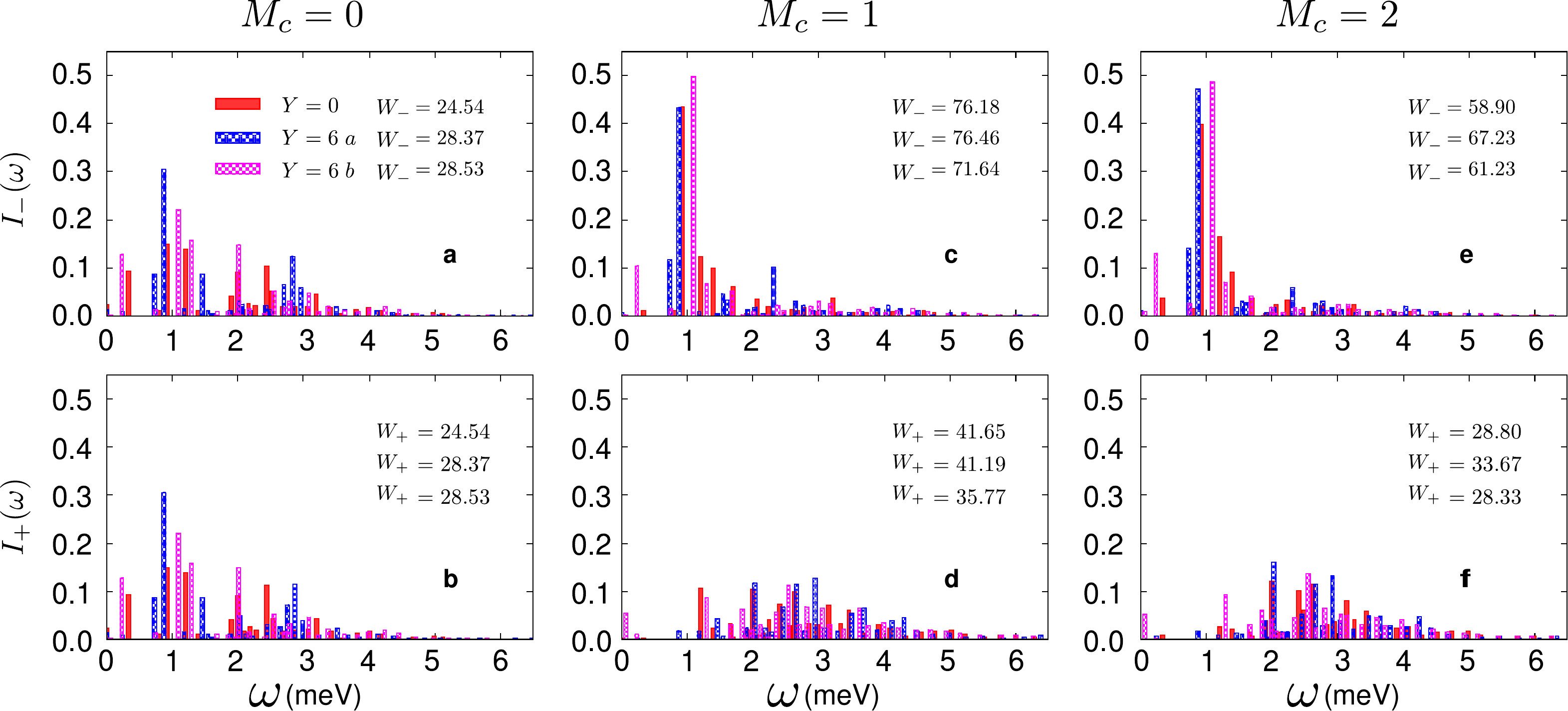}
\caption{\textbf{Chiral graviton spectral functions for the $\boldsymbol{\nu=-9/2}$ state.} The three columns correspond to the states in Fig.~\ref{Fig:Theory1}. The top row is for negative chirality and the bottom row is for positive chirality. In each panel, different symbols are used to represent the three ground states as shown in panel \textbf{a}. For total momentum $Y=6$, the lowest energy state is called $a$ and the second lowest state is called $b$. The three numbers in each panel are the total weights of the spectral functions.}
\label{Fig:Theory2}
\end{figure}

Let us consider the many-body state at $\nu=-9/2$ that occurs without the displacement field. The magnetic field is chosen to be $14$ T so we have $\ell_{B}=6.842$ nm, $d=8.697\ell_{B}$ and $\frac{e^{2}}{4\pi\varepsilon_{0}\varepsilon^{\parallel}_{\rm BN}\ell_{B}} = 31.86$ meV. The LL diagram in Extended Data Fig.~6a is computed using the parameters listed above with $\Delta_{2}=0$ meV. It is helpful to begin our analysis with the $\nu=-6$ state. If the interaction is turned off, electrons fill all the LLs below the $\mathbf{K}_{-},N_{\rm B}=0$ level in Extended Data Fig.~6a. The presence of interaction is not expected to fundamentally change the $\nu=-6$ state since the single-particle gap below $\mathbf{K}_{-},N_{\rm B}=0$ is about $30$ meV. The $\nu=-9/2$ state has an additional $3/2$ filling of electrons compared to the $\nu=-6$ state. The most likely configuration is that electrons are spin-valley polarized and fill the $\mathbf{K}_{-},N_{\rm B}=0$ and $\mathbf{K}_{-},N_{\rm B}=1$ levels. It is essential to keep both levels given their separation is only $4.59$ meV. The energy spectra obtained by exact diagonalization for the system with $36$ electrons and $N_{\phi}=24$ are presented in Fig.~\ref{Fig:Theory1}. Here we only show the low-energy eigenvalues with total momentum $0{\leq}Y{\leq}11$ because the other half is obtained trivially by magnetic translation. In Fig.~\ref{Fig:Theory1}a, there are exactly 24 (12) electrons in the $\mathbf{K}_{-},N_{\rm B}=0$ ($\mathbf{K}_{-},N_{\rm B}=1$) level, hence no mixing is present. In contrast, one or two electrons are allowed to escape from the $\mathbf{K}_{-},N_{\rm B}=0$ level in Fig.~\ref{Fig:Theory1}b,c. For all cases, there are six quasi-degenerate ground states, which is consistent with the prediction for the Moore-Read type states. This information cannot distinguish the Pfaffian, anti-Pfaffian, or particle-hole symmetric Pfaffian states. To this end, we turn to the chiral graviton spectral functions shown in Fig.~\ref{Fig:Theory2}. For the ground states in Fig.~\ref{Fig:Theory1}a, particle-hole symmetry in the $\mathbf{K}_{-},N_{\rm B}=1$ level ensures that the two chiralities have the same the total weights. As soon as LL mixing is turned on, $W_{-}$ becomes much larger than $W_{+}$, which suggests that the ground states in Fig.~\ref{Fig:Theory1}b,c are of the Pfaffian type. In addition, we have also studied the cases with $33$ and $39$ electrons. While ground states cannot be defined for them, their chiral graviton spectral functions also reveal that the negative chirality dominates in the presence of LL mixing. On the experimental side, the weak feature observed at $\nu=-5+7/13$ can be ascribed to the Levin-Halperin daughter state of the Pfaffian state.

Naively, one may expect to obtain an even-denominator FQH state when a NR0/NR1 doublet has $3/2$ filling. This could happen at $\nu=-6+3/2,-4+3/2,-2+3/2,0+3/2$, but only two of them are actually observed. The absence of FQH states at $\nu=-4+3/2$ and $-2+3/2$ is not easy to explain. It may simply because that the sample quality is still insufficient. Another possibility is the competition between spin-valley polarized states and other states that are not fully polarized. The existence of a nonzero $\Delta_{2}$ may reduce some single-particle gaps and/or lead to level crossings. We hope that more in-depth experimental and theoretical studies in the future can shed some light on this issue. While the $\nu=3/2$ state was indeed observed, its nature is more subtle. The active levels are expected to be $\mathbf{K}_{+},N_{\rm B}=0$ and $\mathbf{K}_{+},N_{\rm B}=1$, but the $\mathbf{K}_{+},N_{\rm M}=0$ level is not too far in energy. For our choice of SWMc parameters, the two single-particle gaps between these three levels are $5.22$ meV and $14.8$ meV. The two gaps can be brought to comparable values by increasing the magnetic field and/or changing the SWMc parameters. The fate of the Pfaffian state in this process is unclear but very interesting.

\begin{figure}[ht]
\includegraphics[width=0.80\textwidth]{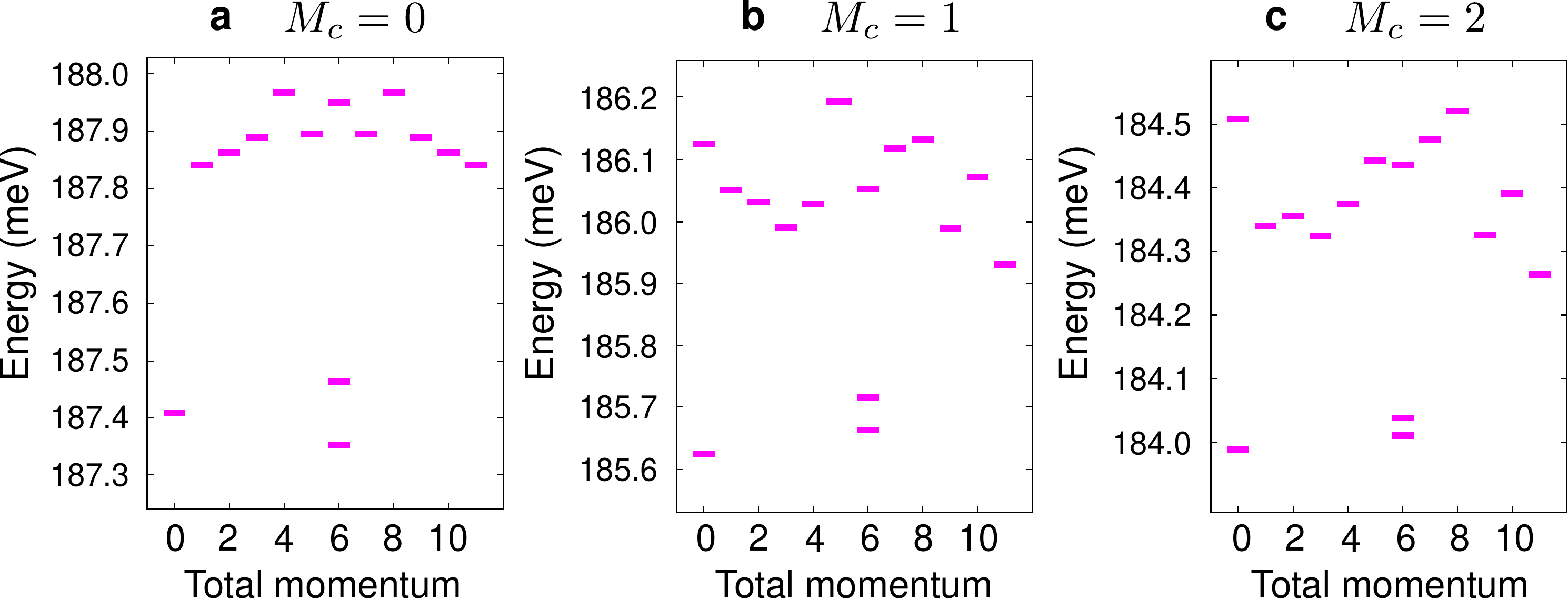}
\caption{\textbf{Low-energy eigenvalues of the $\boldsymbol{\nu=9/2}$ state.} There are 36 electrons in total and the number of electrons that is allowed to escape from the NR0 level is denoted as $M_{c}$.}
\label{Fig:Theory3}
\end{figure}

\begin{figure}[ht]
\includegraphics[width=0.95\textwidth]{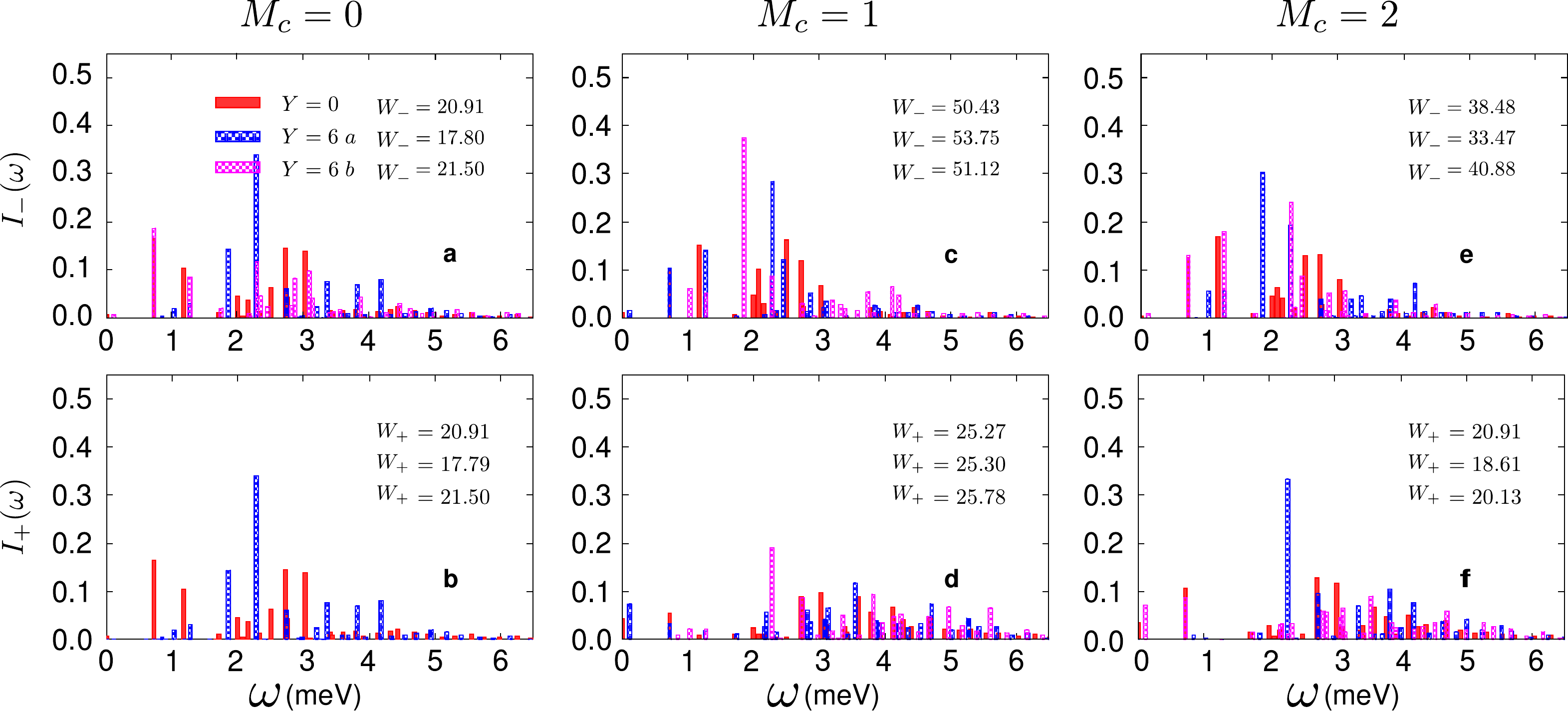}
\caption{\textbf{Chiral graviton spectral functions for the $\boldsymbol{\nu=9/2}$ state.} The three columns correspond to the states in Fig.~\ref{Fig:Theory3}. The symbols are the same as in Fig.~\ref{Fig:Theory2}.}
\label{Fig:Theory4}
\end{figure}

The FQH states at $\nu=-3/2$ and $9/2$ are more difficult to pin down quantitatively. The application of a displacement field results in many LL crossings that is further complicated by a nonzero $\Delta_{2}$. If the electric field is chosen to be $D=155$ mV/nm, we would have $\Delta_{1}=17.31$ meV. The evolution of LLs with $\Delta_{2}$ is shown in Extended Data Fig.~6c. For $\Delta_{2}=0$ meV, the active levels at $\nu=-3/2$ and $9/2$ are of the NR0 type. If $\Delta_{2}=0$ meV is fixed but $D$ increases to a very large value or $D$ is fixed but $\Delta_{2}$ increases to $\sim 5$ meV, one NR1 level in the $\mathbf{K}_{-}$ goes up and crosses with one NR0 level in $\mathbf{K}_{+}$. This enables a Pfaffian type state at $\nu=-3/2$ but the $\nu=9/2$ state is still mysterious. If $\Delta_{2}$ is changed to -10 meV, the top most level (in the $-6{\leq}\nu{\leq}6$ range) would be of the NR1 type. In this case, the $\nu=9/2$ state corresponds to half filling of one spin component of this level, and its mixing with the NR0 type level below it would select the Pfaffian state, as confirmed by the results in Fig.~\ref{Fig:Theory3} and Fig.~\ref{Fig:Theory4}. If we go beyond the one-component spin-valley polarized states, other even-denominator FQH states could be possible. Two examples are the Halperin 331 state~\cite{halperin1983} and the Jain state constructed from parton theory~\cite{jain1989parton}. In this scenario, there is no need to invoke $\Delta_{2}$, but our analysis is still applicable for a small nonzero $\Delta_{2}$. As shown in Extended Data Fig.~6b, two NR0 type levels evolve with $\Delta_{1}$ and become quasi-degenerate in a suitable range. After incorporating the spin degree of freedom, there are four levels whose total filling factor is $5/2$. For the integer part $2$, electrons could populate the same spin but different valleys or the same valley but different spins. The fractional part $1/2$ corresponds to half filling of two NR0 type levels with the same spin or valley index. To realize the Halperin or Jain state, the interlevel interaction should be weaker compared to the intralevel interaction. If the two levels belong to the same valley but have opposite spins, it is difficult to envisage spin-dependent that would favor the Halperin or Jain state. In contrast, the interaction between different valleys may be altered by valley anisotropic terms that arise from lattice scale corrections to the Coulomb potential~\cite{kharitonov2012}. It has been proposed that the Jain state could be realized using suitable valley anisotropic terms~\cite{wu2022two}. To check if this mechanism also works for the TLG, a detailed analysis of valley anisotropic terms would be necessary. This is an interesting topic that is left for future works.

\end{document}